\documentclass[conference]{IEEEtran}
\IEEEoverridecommandlockouts
\usepackage{cite}
\usepackage{amsmath,amssymb,amsfonts}
\usepackage{algorithm}
\usepackage{hyperref}

\usepackage{algpseudocode}
\usepackage{graphicx}
\usepackage{textcomp}
\usepackage{xcolor}
\usepackage[utf8]{inputenc}
\usepackage[T1]{fontenc}
\usepackage[american]{babel}
\usepackage[center]{caption}
\usepackage{graphicx}
\usepackage{listings}
\usepackage{float}
\usepackage{amsmath,amssymb,exscale}
\usepackage{blindtext, graphicx}
\usepackage{verbatim}
\usepackage{algorithm}
\usepackage{algpseudocode}
\usepackage{fancyvrb}
\usepackage{bera}
\usepackage{amsmath}
\usepackage{mathtools}
\usepackage{lipsum}
\usepackage{array}
\usepackage{stfloats}
\usepackage{subcaption}
\usepackage{multirow}
\usepackage{multicol}
\usepackage{scalerel}
\usepackage{tikz}
\usetikzlibrary{svg.path}

\def\BibTeX{{\rm B\kern-.05em{\sc i\kern-.025em b}\kern-.08em
    T\kern-.1667em\lower.7ex\hbox{E}\kern-.125emX}}
\begin{document}

\title{Reliability of Large Language Models for Design Synthesis: An Empirical Study of Variance, Prompt Sensitivity, and Method Scaffolding\\

}
\author{\IEEEauthorblockN{Rabia Iftikhar}
\IEEEauthorblockA{\textit{Institute for Software and Systems Engineering} \\
\textit{Technische Universitat, Clausthal}\\
Clausthal-Zellerfeld, Germany \\
rabiaiftikhar159@gmail.com}
\and
\IEEEauthorblockN{Andreas Rausch}
\IEEEauthorblockA{\textit{Institute for Software and Systems Engineering} \\
\textit{Technische Universitat, Clausthal}\\
Clausthal-Zellerfeld, Germany \\
andreas.rausch@tu-clausthal.de}
}

\maketitle
\begin{abstract}
Large Language Models (LLMs) are increasingly applied to automate software engineering tasks, including the generation of UML class diagrams from natural language descriptions. While prior work demonstrates that LLMs can produce syntactically valid diagrams, syntactic correctness alone does not guarantee meaningful design. This study investigates whether LLMs can move beyond diagram translation to perform design synthesis, and how reliably they maintain design-oriented reasoning under variation. We introduce a preference-based few-shot prompting approach that biases LLM outputs toward designs satisfying object-oriented principles and pattern-consistent structures. Two design-intent benchmarks, each with three domain-only, paraphrased prompts and 10 repeated runs, are used to evaluate three LLMs (ChatGPT 4o-mini, Claude 3.5 Sonnet, Gemini 2.5 Flash) across three modeling strategies: standard prompting, rule-injection prompting, and preference-based prompting, totaling 540 experiments (i.e. 2x3x10x3x3). Results indicate that while preference-based alignment improves adherence to design intent it does not eliminate non-determinism, and model-level behavior strongly influences design reliability. These findings highlight that achieving dependable LLM-assisted software design requires not only effective prompting but also careful consideration of model behavior and robustness.

\end{abstract}

\footnote{For the reproducibility of the experiments conducted in this research, the data is available at \cite{repo}}


\section{Introduction}
Large Language Models (LLMs) are increasingly used to automate software engineering tasks \cite{four,six,five,promptPattern}, including the generation of design artifacts such as UML class diagrams \cite{behav,student}. Recent tools and studies show that LLMs can produce syntactically valid diagrams from natural language descriptions, however, syntactic validity alone does not constitute good software design that ensures maintainability, extensibility, and architectural soundness. In practice, many LLM-generated diagrams remain structurally shallow: they mirror surface-level entities mentioned in text but fail to capture the design intent: the design principles (i.e. abstraction, encapsulation, separation of concern), and design knowledge (i.e. GoF design patterns) the experienced designers apply to promote extensibility and reusability \cite{twelve,behav,student}. When this knowledge is absent, the modeling process risks becoming "garbage in, garbage out,” where small changes or omissions in the input lead to inconsistent diagrams that cannot be trusted for downstream implementation \cite{five,nine,trust}.

This limitation highlights a fundamental distinction between diagram translation and design synthesis. Many current AI-based modeling approaches behave primarily as translators, mapping textual nouns and verbs to classes, attributes, and associations. Yet high-quality architectural modeling requires more than element extraction—it requires inferring implicit design patterns and principles from requirements. Importantly, real-world requirements rarely specify which design pattern or principles to apply. Instead, they describe domain rules and invariants (e.g., dynamic user roles or policy-dependent behavior), from which engineers derive appropriate architectural structures. Therefore, an LLM-based modeling assistant should be evaluated on whether it can implicitly infer and apply design principles and knowledge rather than merely reproducing patterns and principles named in prompts \cite{wei2022chain,bubeck2023sparks}.

Beyond design quality, an equally critical but underexplored challenge is behavioral reliability. Generative models are inherently non-deterministic, yet architectural workflows demand stability and repeatability. Practitioners must be able to expect structurally consistent designs under realistic variations such as prompt paraphrasing or repeated executions. If outputs fluctuate significantly, their practical utility diminishes regardless of average performance. Despite growing adoption of LLM-assisted development, limited empirical work has systematically examined how reliably frontier models sustain design-level reasoning \cite{pearce2023asleep}. Existing studies largely emphasize capability while overlooking behavioral stability across runs, sensitivity to prompt variation, and differences across model families \cite{bubeck2023sparks, pearce2023asleep,wang2023selfconsistency}. For architecture-centric tasks, where inconsistencies may propagate into implementation defects, this omission is particularly consequential. 

This leads to the following research questions in this study. \textbf{RQ1:} To what extent can LLMs generate design-level UML class diagrams that satisfy domain constraints and design principles when provided with domain-only descriptions? \textbf{RQ2:} To what extent do LLM-generated diagrams exhibit pattern-consistent structures implicitly, and how does the proposed approach affect the emergence of such structures compared to standard prompting and rule-injection baselines? \textbf{RQ3:} How robust are the generated designs across repeated runs and prompt paraphrases, and what behavioral stability differences emerge across models? \textbf{RQ4:} How does increasing domain complexity affect adherence to design principles, pattern emergence, and behavioral stability?

This paper investigates not only whether LLM behavior can be guided from translation toward design synthesis, but also how reliably different models maintain design-oriented reasoning under controlled sources of variation. We propose a preference-based few-shot prompting approach that exposes models to contrasting modeling solutions and biases generation toward designs that satisfy object-oriented principles and pattern-consistent structures \footnote{A preliminary version of the preference-based alignment idea was previously presented as an abstract-only contribution at a workshop venue \cite{Iftikhar2025}; the present paper substantially extends that early concept by providing a concrete implementation, a controlled experimental protocol, and a comprehensive empirical evaluation across multiple models, scenarios, and sources of variation.}. To avoid trivializing the task we construct two varied domain, design-intent benchmarks in which textual problem statements describe domain behavior and constraints without naming principles and patterns. For each design task, we compare three LLMs (i.e. ChatGPT 4o-mini, Claude 3.5 Sonnet, Gemini 2.5 Flash), on three paraphrased prompts, using three different modeling approaches (i) standard prompting baselines, (ii) a rule-injection baseline that explicitly encodes design principles, and (iii) preference-based prompting, for 10 repeated runs (i.e. 2x3x3x3x10=540 runs). 

From our results a clear pattern emerges: performance differences cannot be attributed solely to prompting strategy when applied to design synthesis tasks. Instead, model-level behavioral characteristics strongly influence reliability in design synthesis tasks. While the proposed preference-based alignment improves the likelihood of generating models that reflect architectural intent, we observe distinct stability regimes across the evaluated models—Claude exhibiting high decoding stability across repeated runs, ChatGPT demonstrating predictable behavior once design scaffolding is defined, and Gemini showing substantial variability across paraphrases and executions. These findings suggest that model choice may be a dominant factor in achieving dependable software designs, reframing LLM-based modeling as not only a capability problem but also a reliability challenge.

\textit{Paper Structure:} The remainder of the paper is structured as follows: Section 2 reviews related work. Section 3 outlines the experimental design. Section 4 describes the technical implementation of the prompting strategies used within that protocol. Section 5 reports evaluation results and discusses findings with respect to the research questions. Section 6 outlines study limitations, and section 7 concludes the paper and identifies directions for future research.

\section{Background and Related Work}
\textbf{LLMs for Software Design}
Early studies demonstrated that models like Codex and GPT-3 can generate syntactically correct code from natural language prompts, highlighting their potential as intelligent programming assistants \cite{chen2021evaluating}. Subsequent work extended these capabilities to software design, investigating whether LLMs can produce UML diagrams from textual descriptions \cite{Taglia,behav,ADD,student}, suggesting that natural language-based design automation may support early-stage modeling activities.

Despite these promising capabilities, researchers caution that syntactic correctness does not guarantee design quality. LLM-generated artifacts often reflect surface-level textual correlations rather than underlying architectural reasoning \cite{pearce2023asleep, wang2023selfconsistency}. This observation aligns with broader concerns about hallucinations in generative AI, where outputs are plausible but semantically inconsistent or incomplete relative to domain requirements \cite{bubeck2023sparks, student,ojha2025}.

\textbf{Design Quality and Pattern Consistency}
Studies examining LLM-assisted modeling have shown that models frequently fail to internalize design knowledge, producing diagrams that lack meaningful class hierarchies, appropriate encapsulation, or pattern-consistent structures \cite{wei2022chain, student, behav}. For instance, \cite{pearce2023asleep} highlight that while LLMs can identify classes and relationships from text, they often misrepresent associations or omit key behavioral constraints. This limitation underlines the distinction between diagram translation and true design synthesis.

\textbf{Prompt Engineering and Preference-based Alignment}
Recent efforts attempt to bridge this gap using prompt engineering techniques, including chain-of-thought reasoning and few-shot learning, to scaffold model reasoning aligning it with object-oriented principles \cite{wei2022chain, L8, twelve}. Other approaches incorporate explicit rule-based knowledge or knowledge injection to improve the adherence of generated diagrams to design principles \cite{L1, twelve}. However, these strategies often remain partial solutions, as they depend heavily on domain modeling or domain-specific rules, limiting generalization to unseen design scenarios.

Another line of work moves beyond translation and toward design synthesis using preference-based alignment and reinforcement learning from human feedback (RLHF). These methods expose models to contrasting solutions and encourage preference for outputs that satisfy architectural principles, pattern consistency, and maintainability considerations \cite{twenty}. Applied to UML diagram generation, such approaches can bias LLM outputs toward designs that reflect latent domain knowledge without requiring explicit pattern names, providing a pathway for more robust, design-oriented modeling.

\textbf{Behavioral Reliability and Model Variability}
Beyond correctness, LLMs introduce challenges in behavioral reliability. Generative models are inherently stochastic, and outputs may vary across repeated runs, prompt paraphrases, or model families \cite{wang2023selfconsistency, pearce2023asleep} undermining trust in AI-assisted workflows \cite{trust, L7}. Comparative studies indicate that different models exhibit distinct stability regimes, with some models producing highly repeatable outputs under structured prompting, while others fluctuate across executions \cite{bubeck2023sparks, wang2023selfconsistency}. These observations suggest that evaluating LLMs for design synthesis requires not only assessing the fidelity of generated artifacts but also their reliability under realistic variations such as repeated runs, prompt paraphrasing, or cross-model comparisons.

\textbf{Perspective of current study:}
In summary, prior research demonstrates that while LLMs are capable of producing syntactically valid diagrams, achieving design synthesis requires addressing three intertwined challenges: internalization of design principles, behavioral stability across variations, and evaluation frameworks that capture both correctness and reliability. Our study builds on these insights, systematically investigating how model choice, and prompting strategy interact to influence both the quality and robustness of generated UML diagrams.

\section{Experimental Design}
We evaluate three state-of-the-art models representing different architectures and decoding characteristics (i.e. ChatGPT 4o-mini, Gemini 2.5 Flash, Claude 3.5 Sonnet). These models are evaluated under three different approaches (i) standard prompting, (ii) rule-injected prompting, (iii) preference-based few-shot prompting, with three paraphrased prompts per benchmark, and a temperature value set to 0.5. 
\subsection{Experiment Conduction}
To ensure realistic evaluation, we constructed two domain-focused benchmarks. A medium-complexity system describing policy-based behavior in a hospital billing system (HMS), and a high-complexity system modeling sensor networks with event-driven interactions. For each benchmark, prompts are phrased thrice without naming any design patterns or rules, ensuring that models must infer architectural structures implicitly.

Per benchmark, we ran 10 experiments per paraphrased prompt with each model (ChatGPT, Claude, Gemini) and each prompting methodology (standard, rule-injected, preference-based) collecting a total of 540 UML diagrams: $2 \text{ domains} \times 3 \text{ prompts} \times 3 \text{ models} \times 3 \text{ prompting approaches} \times 10 \text{ runs} = 540$ outputs. We then evaluated each diagram for correctness, principles and pattern consistency, and stability against the reference model designed by an MDE focused Ph.D. student for each benchmark. To reduce evaluator bias, the reference model and rule mapping were independently reviewed by an additional modeling expert with industrial experience and the disagreements were resolved through discussion. From the reference models of these benchmarks we identified two design principles (i.e. abstraction and encapsulation), and two design patterns (i.e. Observer pattern, strategy pattern) in total. Table \ref{tab:Principles-in-reference-model} lists these principles and patterns, their literature sources, and how they are instantiated in each reference model. 

\begin{table*}
    \centering
    \scriptsize
    \begin{tabular}{|c|c|c|}
    \hline
    \textbf{Principles and Patterns} & \textbf{Application in HMS Reference Model} & \textbf{Application in Sensor Reference Model}\\
\hline
    Abstraction\cite{two, sixteen} & Patient, InPatient, OutPatient class& \tiny Sensor, PressureSensor, TemperatureSensor, RainGauge, WindSensor \\ 
\hline
Encapsulation\cite{two, sixteen}&   Private attributes used & Private attributes used\\ 
\hline
Observer Pattern\cite{designpattern, hunt1999_pragmatic_programmer}& Not Applicable &\tiny Subject, Observer, WeatherInformationSystem, WeatherStation \\ 
\hline
Strategy Pattern\cite{designpattern} & \tiny BillingStrategy, BillingProcessor,InsuranceBilling, SelfPayBilling, GovernmentBillings &\tiny RepotingStrategy, ReportingContext, HourlyReporting, CustomIntervalReporting\\
\hline
    
    \end{tabular}
    \caption{Application of established design principles and patterns in Reference Models}
    \label{tab:Principles-in-reference-model}
\end{table*}

\subsection{Evaluation Settings}
We conduct a manual evaluation to better capture the architectural quality of generated models. This decision is motivated by recent work in LLM-based code and model generation, which just not only highlight the shortcomings of automated evaluation of the models\cite{twelve} but also token-level metrics for measuring code or model generation quality \cite{frontier2025} emphasizing the need for semantic, structural, and expert‑informed evaluation methods that better reflect real design quality and usability \cite{camara2023, camara2024benchmarks}.
\subsubsection{Evaluation Metrics}
For each model, we define three metrics:
\begin{itemize}
    \item \textbf{M1.1: Structural Correctness Score:} Measures the degree to which the structural elements in the generated diagram match expert reference model.\\
    \textit{Computation:} Let $TP$ be the number of correctly predicted elements, $FP$ false positives, and $FN$ false negatives. Then:
\[
Precision = \frac{TP}{TP + FP}\]
\\
\[\quad Recall = \frac{TP}{TP + FN}\] \\
\[\quad F1 = \frac{2 \cdot Precision \cdot Recall}{Precision + Recall}\]
However, this metric is token-level and does not fully reflect design intent of the generated model. Hence, it is supplemented by semantic metrics defined below.

\item \textbf{M1.2: Principle Adherence Score:} Measures the extent to which the diagram adheres to established OOAD principles identified in reference model.\\
\textit{Computation:} A separate checklist of design rules is used for each benchmark (see table. \ref{tab:Principles-in-reference-model}). Each diagram is assessed for adherence to each rule:
\[
\text{Principle Adherence Score} = \frac{\# \text{Rules Satisfied}}{\text{Total Rules Evaluated}}
\]
This assessment is conducted by the authors of this paper using the reference model and associated rule mapping on ternary scale (0=no, 0.5= partially, 1=yes).

\item \textbf{M1.3: Pattern Emergence Score:} Measures whether the generated diagram instantiate a specific design pattern identified in reference model.\\
Each diagram is assessed for emergence of design pattern:
\[
\text{Pattern Emergence Score} = \frac{\# \text{Pattern Applied}}{\text{Total Patterns Evaluated}}
\]
This assessment is conducted by the authors of this paper using the reference model and associated rule mapping on ternary scale (0=no, 0.5= partially, 1=yes).

\item \textbf{M1.4: Stability Index (SI):} Measures variance in diagrams generated across runs and prompt variations, capturing behavioral reliability. Higher SI indicates better consistency.

\textit{Computation:}
\[
\text{SI} = \frac{1}{1 + \left( \sigma_{\text{prompt}}^2 + \sigma_{\text{run}}^2 \right)}
\]
\end{itemize}
To ensure systematic alignment between research questions, evaluation criteria, and metrics, each RQ is explicitly operationalized as shown in table \ref{tab:rq_metric_mapping}.
\begin{table}[t]
\centering
\begin{tabular}{|l|l|}
\hline
\textbf{RQ} & \textbf{Evaluation Metric} \\
\hline 
RQ1 & M1.2. Principle Adherence Score (PAS) \\
\hline
RQ2 & M1.3. Pattern Emergence Score (PES) \\
\hline
RQ3 & M1.4. Stability Index (SI) \\
\hline
RQ4 & Comparative analysis of PAS, PES, and SI \\
\hline

\end{tabular}
\caption{Mapping of RQs to Evaluation Metrics}
\label{tab:rq_metric_mapping}

\end{table}

\section{Prompting Methods}
To understand the limitations of LLMs in generating design-level UML class diagram, we conduct a series of experiments using three different prompting methods. 
\subsection{Standard Baseline: Standard LLM Prompting}
In this experimental condition, we evaluate the raw design-generation capability of the selected LLMs by providing only the domain-level design task within the prompt, without any additional rule injection or role based prompting. The objective of this setup is to observe how the models naturally interpret and respond to architectural modeling tasks when operating under minimal instruction. By isolating the task description from external constraints or injected design knowledge, this condition serves as a baseline for assessing whether LLMs inherently perform design synthesis or merely translate textual entities into structural elements. This configuration allows us to characterize the default modeling behavior of each LLM and establish a reference point for comparison against rule-injection and proposed preference-based prompting approaches.
\subsection{Rule-injected Baseline: Prompting with Rule Injection}
The purpose of this experiment was to analyze to what extent does the use of structured prompts with explicitly injected general design principles and expert advices can guide the LLMs to generate design-level UML class diagrams. Unlike most previous work, which implicitly assumes LLMs understand good design practices, this setup deliberately encodes those expectations into the prompt. This experiment has been done as a necessary control to test whether prior LLMs under performed due to lack of architectural exposure, rather than capability. The prompt passed to each model was divided into three structural components: Priming (get the LLM ready using role-play prompting and embed the established OOAD design principles derived from literature), Question (the core class diagram generation task), and Decorator (the expert guidance on formatting, relationship modeling, and a self-check phase)
\subsection{Proposed Approach: Preference-based Few-shot Prompting}
In this section we present a preference-guided few-shot prompting approach. While our rule-injected baseline incorporated design rule prompts, this section introduces an additional guidance strategy, that in addition to design rule prompts, trains the LLM to recognize better and worse design alternatives through annotated preference-based examples. Of particular concern, it is not a full alignment pipeline like RLHF, instead it is a lightweight, empirical technique intended to explore whether preference signals shift the LLMs output in design domain towards architecturally sound structures.

\begin{figure}
    \centering
    \includegraphics[width=1\linewidth]{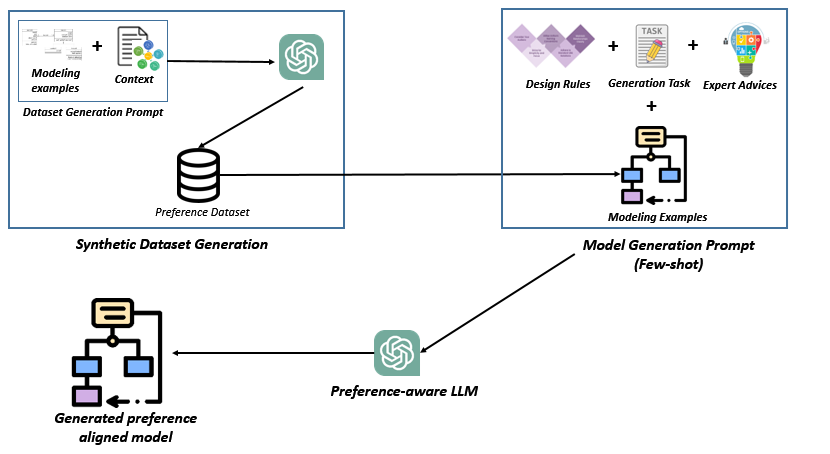}
    \caption{Proposed Approach}
    \label{fig:proposed-approach}
\end{figure}

\subsubsection{Preference Dataset Generation}
To generate our modeling dataset we carry out a whole new set of conversation with LLM (o1-mini) on good design principles and patterns. The aim of conducting this conversation was to provide enough modeling context to the LLM before it generates a cross-domain modeling dataset consisting of a set of five task description and a design solution. Since LLMs are not good at design task, the generated diagram was also not up-to the mark and hence marked as a bad solution. The idea behind this decision was to let the LLM know that the way it generates diagram is not a preferred human choice. For a good design solution, we asked our modeling experts to define good design solutions against the generated problems. To be clear, all five data points in the dataset include a design-level description, two generated class diagram solutions (solution1: by LLM, solution2: by modeling experts), and preferred choice labeled by a human annotator (i.e. the expert generated model). The preference was manually annotated by the authors of this paper based on established OOAD principles and then validated by one modeling expert from industry. 

It is to be noted that the benchmarks used in this study (i.e. HMS and Sensor Network) to evaluate the LLMs behavioral stability and reliability for design synthesis are not part of this dataset. Those benchmarks only contain the design description and one expert designed reference model to facilitate evaluation. 

\section{Results and Discussion}
This section presents the findings of the study organized around the research questions. Rather than evaluating design generation purely as a capability problem, the analysis focuses on design quality, behavioral stability, and the conditions under which LLMs can reliably support architectural modeling.

\subsection{Answer to RQ1}

Across both domains and all evaluated models, the preference-aware approach consistently produced diagrams that more closely aligned with the reference architectures. Models operating under this setting applied a greater proportion of expected object-oriented principles (see table \ref{tab:PASpref}). These results suggest that preference signals can guide models beyond surface-level entity extraction toward more deliberate architectural organization.

Among the three models, GPT demonstrated the strongest adherence to design principles in the preference-based setting, applying the highest number of architectural constructs present in the reference models, however, occasionally produced duplicate inheritance relationships and demonstrated partial constraint satisfaction. Claude generated structurally coherent diagrams but failed to abstract shared behavior into superclasses. Gemini showed comparatively weaker performance, often messing with key structural elements required to satisfy domain constraints including duplicate and circular relationships, conflicting inheritance hierarchies, and missing classes.

Minimal differences were observed between standard prompting (see table \ref{tab:PASstandard}) and rule injection for most models (see table \ref{tab:PASfair}), indicating that explicitly encoding design rules does not necessarily translate into improved architectural reasoning. Notably, Gemini’s performance degraded under rule injection, potentially due to the increased cognitive load imposed by dense instructions, which appeared to trigger hallucinated relationships and inconsistent structures.

\textit{Key Insight:} LLMs may generate design-level diagrams when guided appropriately, but preference-based prompting appears more effective than prescriptive rule injection in supporting architectural reasoning.
\begin{table*}
\centering
\scriptsize
\begin{tabular}{|c|c|ccc|ccc|ccc|}
\hline

\multirow{2}{*}{\textbf{Model}} & \multirow{2}{*}{\textbf{Prompt}}
& \multicolumn{3}{|c|}{\textbf{Standard}}
& \multicolumn{3}{|c|}{\textbf{Rule-injected}}
& \multicolumn{3}{|c|}{\textbf{Preference}} \\
\cline{3-11}
& & \textbf{Pr} & \textbf{Re} & \textbf{F1} & \textbf{Pr} & \textbf{Re} & \textbf{F1} & \textbf{Pr} & \textbf{Re} & \textbf{F1} \\
\hline

Claude & P1 & 0.3333 & 0.667 & 0.4444 & 0.3636 & 0.667 & 0.4706 & 0.4444 & 0.6667 & 0.5333 \\
Claude & P2 & 0.3333 & 0.667 & 0.4444 & 0.3333 & 0.667 & 0.4444 & 0.4 & 0.6667 & 0.5 \\
Claude & P3 & 0 & 0 & 0 & 0.3636 & 0.667 & 0.4706 & 0.4 & 0.6667 & 0.5 \\
\hline

GPT & P1 & 0.3 & 0.6 & 0.4 & 0.3636 & 0.6667 & 0.4706 & 0.2353 & 0.6667 & 0.3478 \\
GPT & P2 & 0.3 & 0.6 & 0.4 & 0.3636 & 0.6667 & 0.4706 & 0.2353 & 0.6667 & 0.3478 \\
GPT & P3 & 0.3 & 0.6 & 0.4 & 0.3636 & 0.6667 & 0.4706 & 0.2353 & 0.6667 & 0.3478 \\
\hline

Gemini & P1 & 0.0254 & 0.5 & 0.0484 & 0.2632 & 0.625 & 0.3704 & 0.2727 & 0.5 & 0.3529 \\
Gemini & P2 & 0.0266 & 0.5 & 0.0504 & 0.1933 & 0.5088 & 0.2802 & 0.2703 & 0.5 & 0.3509 \\
Gemini & P3 & 0.0242 & 0.5 & 0.0462 & 0.1915 & 0.4737 & 0.2727 & 0.2703 & 0.5 & 0.3509 \\
\hline
\end{tabular}
\caption{M1.1: HMS-Concrete Class Performance Averaged across 10 Runs}
\label{tab:HMS-cc}
\end{table*}

\begin{table*}[htbp]
\centering
\scriptsize
\begin{tabular}{|c|c|ccc|ccc|ccc|}
\hline
\multirow{2}{*}{\textbf{Model}} & \multirow{2}{*}{\textbf{Prompt}}
& \multicolumn{3}{|c|}{\textbf{Standard}}
& \multicolumn{3}{|c|}{\textbf{Rule-injected}}
& \multicolumn{3}{|c|}{\textbf{Preference}} \\
\cline{3-11}
& & \textbf{Pr} & \textbf{Re} & \textbf{F1} & \textbf{Pr} & \textbf{Re} & \textbf{F1} & \textbf{Pr} & \textbf{Re} & \textbf{F1} \\
\hline
Claude & P1 & 0 & 0 & 0 & 0 & 0 & 0 & 0.5 & 0.5 & 0.5 \\
Claude & P2 & 0 & 0 & 0 & 0 & 0 & 0 & 0.5 & 0.5 & 0.5 \\
Claude & P3 & 0 & 0 & 0 & 0 & 0 & 0 & 0.5 & 0.5 & 0.5 \\
\hline

GPT & P1 & 0 & 0 & 0 & 0 & 0 & 0 & 0.3333 & 0.5 & 0.4 \\
GPT & P2 & 0 & 0 & 0 & 0 & 0 & 0 & 0.3333 & 0.5 & 0.4 \\
GPT & P3 & 0 & 0 & 0 & 0 & 0 & 0 & 0.3333 & 0.5 & 0.4 \\
\hline

Gemini & P1 & 0.0099 & 0.0556 & 0.0168 & 0.0909 & 0.4 & 0.1482 & 0.2 & 0.125 & 0.1539 \\
Gemini & P2 & 0 & 0 & 0 & 0 & 0 & 0 & 0.0909 & 0.05 & 0.0645 \\
Gemini & P3 & 0.0099 & 0.0556 & 0.0168 & 0 & 0 & 0 & 0.0909 & 0.05 & 0.0645 \\
\hline
\end{tabular}
\caption{M1.1: HMS-Abstract Class Performance Averaged across 10 Runs}
\label{tab:HMSac}
\end{table*}

\begin{table*}[htbp]
\centering
\scriptsize
\begin{tabular}{|c|c|ccc|ccc|ccc|}
\hline
\multirow{2}{*}{\textbf{Model}} & \multirow{2}{*}{\textbf{Prompt}}
& \multicolumn{3}{|c|}{\textbf{Standard}}
& \multicolumn{3}{|c|}{\textbf{Rule-injected}}
& \multicolumn{3}{|c|}{\textbf{Preference}} \\
\cline{3-11}
& & \textbf{Pr} & \textbf{Re} & \textbf{F1} & \textbf{Pr} & \textbf{Re} & \textbf{F1} & \textbf{Pr} & \textbf{Re} & \textbf{F1} \\
\hline
Claude & P1 & 0 & 0 & 0 & 0 & 0 & 0 & 0.4 & 0.5714 & 0.4706 \\
Claude & P2 & 0 & 0 & 0 & 0.0667 & 0.1429 & 0.0909 & 0.2727 & 0.4286 & 0.3333 \\
Claude & P3 & 0 & 0 & 0 & 0 & 0 & 0 & 0.3636 & 0.5714 & 0.4444 \\
\hline

GPT & P1 & 0 & 0 & 0 & 0.0833 & 0.1429 & 0.1053 & 0.15 & 0.4286 & 0.2222 \\
GPT & P2 & 0 & 0 & 0 & 0.0833 & 0.1429 & 0.1053 & 0.15 & 0.4286 & 0.2222 \\
GPT & P3 & 0 & 0 & 0 & 0.0833 & 0.1429 & 0.1053 & 0.15 & 0.4286 & 0.2222 \\
\hline

Gemini & P1 & 0 & 0 & 0 & 0 & 0 & 0 & 0.0566 & 0.2857 & 0.0945 \\
Gemini & P2 & 0.0323 & 0.0536 & 0.0403 & 0 & 0 & 0 & 0.0411 & 0.0429 & 0.0420 \\
Gemini & P3 & 0 & 0 & 0 & 0 & 0 & 0 & 0.0361 & 0.0429 & 0.0392 \\
\hline
\end{tabular}
\caption{M1.1: HMS-Relationship Performance Averaged across 10 runs }
\label{tab:HMSRel}
\end{table*}
\begin{table*}[htbp]
\centering
\scriptsize
\begin{tabular}{|c|c|ccc|ccc|ccc|}
\hline
\multirow{2}{*}{\textbf{Model}} & \multirow{2}{*}{\textbf{Prompt}}
& \multicolumn{3}{|c|}{\textbf{Standard}}
& \multicolumn{3}{|c|}{\textbf{Rule-injected}}
& \multicolumn{3}{|c|}{\textbf{Preference}} \\
\cline{3-11}
& & \textbf{Pr} & \textbf{Re} & \textbf{F1} & \textbf{Pr} & \textbf{Re} & \textbf{F1} & \textbf{Pr} & \textbf{Re} & \textbf{F1} \\
\hline
Claude & P1 & 0.6 & 0.6667 & 0.6316 & 0.0909 & 0.1111 & 0.1 & 0.5 & 0.5556 & 0.5263 \\
Claude & P2 & 0.0909 & 0.1111 & 0.1 & 0.125 & 0.1111 & 0.1176 & 0.3333 & 0.4444 & 0.3810 \\
Claude & P3 & 0.0833 & 0.1111 & 0.0952 & 0.0909 & 0.1111 & 0.1 & 0.4 & 0.4444 & 0.4211 \\
\hline

GPT & P1 & 0.4 & 0.2222 & 0.2857 & 0.5 & 0.2222 & 0.3077 & 0.3333 & 0.1111 & 0.1667 \\
GPT & P2 & 0.4 & 0.2222 & 0.2857 & 0.5 & 0.2222 & 0.3077 & 0.3333 & 0.1111 & 0.1667 \\
GPT & P3 & 0.4 & 0.2222 & 0.2857 & 0.5 & 0.2222 & 0.3077 & 0.3333 & 0.1111 & 0.1667 \\
\hline

Gemini & P1 & 0.3051 & 0.2222 & 0.2571 & 0.2394 & 0.1889 & 0.2112 & 0.2273 & 0.3704 & 0.2817 \\
Gemini & P2 & 0.1 & 0.1111 & 0.1053 & 0.1563 & 0.1111 & 0.1299 & 0.3462 & 0.5143 & 0.4138 \\
Gemini & P3 & 0.2128 & 0.1111 & 0.1460 & 0.1667 & 0.1111 & 0.1333 & 0.3462 & 0.5143 & 0.4138 \\
\hline
\end{tabular}
\caption{M1.1: Sensor Network— Concrete Class Performance Averaged across 10 runs}
\label{tab:WScc}
\end{table*}

\begin{table*}[htbp]
\centering
\scriptsize
\begin{tabular}{|c|c|ccc|ccc|ccc|}
\hline
\multirow{2}{*}{\textbf{Model}} & \multirow{2}{*}{\textbf{Prompt}}
& \multicolumn{3}{|c|}{\textbf{Standard}}
& \multicolumn{3}{|c|}{\textbf{Rule-injected}}
& \multicolumn{3}{|c|}{\textbf{Preference}} \\
\cline{3-11}
& & \textbf{Pr} & \textbf{Re} & \textbf{F1} & \textbf{Pr} & \textbf{Re} & \textbf{F1} & \textbf{Pr} & \textbf{Re} & \textbf{F1} \\
\hline
Claude & P1 & 1 & 0.25 & 0.4 & 0 & 0 & 0 & 0.5 & 0.25 & 0.3333 \\
Claude & P2 & 0 & 0 & 0 & 0 & 0 & 0 & 0.3333 & 0.25 & 0.2857 \\
Claude & P3 & 0 & 0 & 0 & 0 & 0 & 0 & 0.5 & 0.25 & 0.3333 \\
\hline

GPT & P1 & 0 & 0 & 0 & 0 & 0 & 0 & 0 & 0 & 0 \\
GPT & P2 & 0 & 0 & 0 & 0 & 0 & 0 & 0 & 0 & 0 \\
GPT & P3 & 0 & 0 & 0 & 0 & 0 & 0 & 0 & 0 & 0 \\
\hline

Gemini & P1 & 0 & 0 & 0 & 0 & 0 & 0 & 0.5 & 0.025 & 0.0476 \\
Gemini & P2 & 0 & 0 & 0 & 0 & 0 & 0 & 0 & 0 & 0 \\
Gemini & P3 & 0 & 0 & 0 & 0 & 0 & 0 & 0 & 0 & 0 \\
\hline
\end{tabular}
\caption{M1.1: Sensor Netwrok— Abstract Class Performance Averaged across 10 runs}
\label{tab:WSac}
\end{table*}

\begin{table*}[htbp]
\centering
\scriptsize
\begin{tabular}{|c|c|ccc|ccc|ccc|}
\hline
\multirow{2}{*}{\textbf{Model}} & \multirow{2}{*}{\textbf{Prompt}}
& \multicolumn{3}{|c|}{\textbf{Standard}}
& \multicolumn{3}{|c|}{\textbf{Rule-injected}}
& \multicolumn{3}{|c|}{\textbf{Preference}} \\
\cline{3-11}
& & \textbf{Pr} & \textbf{Re} & \textbf{F1} & \textbf{Pr} & \textbf{Re} & \textbf{F1} & \textbf{Pr} & \textbf{Re} & \textbf{F1} \\
\hline
Claude & P1 & 0.2727 & 0.2143 & 0.24 & 0 & 0 & 0 & 0.25 & 0.2308 & 0.24 \\
Claude & P2 & 0 & 0 & 0 & 0 & 0 & 0 & 0.2 & 0.2308 & 0.2143 \\
Claude & P3 & 0 & 0 & 0 & 0 & 0 & 0 & 0.2308 & 0.2308 & 0.2308 \\
\hline

GPT & P1 & 0.25 & 0.0769 & 0.1176 & 0.25 & 0.0769 & 0.1176 & 0 & 0 & 0 \\
GPT & P2 & 0.25 & 0.0769 & 0.1176 & 0.25 & 0.0769 & 0.1176 & 0 & 0 & 0 \\
GPT & P3 & 0.25 & 0.0769 & 0.1176 & 0.25 & 0.0769 & 0.1176 & 0 & 0 & 0 \\
\hline

Gemini & P1 & 0.1389 & 0.0769 & 0.0990 & 0.1346 & 0.3627 & 0.1964 & 0.0024 & 0.0077 & 0.0036 \\
Gemini & P2 & 0 & 0 & 0 & 0 & 0 & 0 & 0 & 0 & 0 \\
Gemini & P3 & 0.2128 & 0.5556 & 0.1460 & 0 & 0 & 0 & 0 & 0 & 0 \\
\hline
\end{tabular}
\caption{M1.1: Sensor Network— Relationship Performance Averaged across 10 runs}
\label{tab:WSRel}
\end{table*}

\subsection{Answer to RQ2}

The preference-based approach increased the likelihood that models would produce pattern-consistent structures without explicitly naming patterns in the prompt (see table \ref{tab:PESstandard}-\ref{tab:PESpref}). This indicates that exposure to contrasting architectural solutions can bias models toward reusable design strategies rather than direct textual translation.

However, recurring design smells reveal that implicit pattern application remains incomplete. Claude frequently omitted aggregation relationships between the Context class and Strategy interface. GPT occasionally produced duplicate inheritance relationships. Gemini exhibited the highest concentration of structural issues, including conflicting inheritance hierarchies, and missing classes.

These findings suggest that while models can approximate pattern structures, they often struggle with enforcing the full set of architectural constraints required for correct implementation.

\textit{Key Insight:} Preference optimization encourages pattern emergence, but current models still lack the global reasoning necessary to consistently operationalize design patterns.

\begin{table*}[t]
\centering
\scriptsize

\begin{tabular}{|c|c|ccc|ccc|ccc|}
\hline

\multirow{2}{*}{\textbf{Task}} & \multirow{2}{*}{\textbf{Principle}}
& \multicolumn{3}{|c|}{\textbf{ChatGPT}}
& \multicolumn{3}{|c|}{\textbf{Gemini}}
& \multicolumn{3}{|c|}{\textbf{Claude}}\\

\cline{3-11}

& & P1 & P2 & P3 & P1 & P2 & P3 & P1 & P2 & P3  \\

\hline
\multirow{3}{*}{\textbf{HMS}}
& Abstraction        & 0 & 0 & 0 & 0 & 0 & 0 & 0 & 0 & 0\\ \cline{2-11}
& Encapsulation      & 0 & 0 & 0 & 1 & 1 & 1 & 1 & 0 & 1  \\ \cline{2-11}

\hline
\multirow{4}{*}{\textbf{Sensor Netwrok}}
& Abstraction        & 0 & 0 & 0 & 0 & 0 & 0 & 1 & 0 & 0 \\ \cline{2-11}
& Encapsulation      & 0 & 0 & 0 & 0.5 & 0.5 & 0.5 & 1 & 1 & 1 \\ \cline{2-11}

\hline
\end{tabular}

\caption{M1.2: Principle Adherence Score — Standard Baseline Averaged across 10 runs}
\label{tab:PASstandard}
\end{table*}
\begin{table*}[t]
\centering
\scriptsize
\begin{tabular}{|c|c|ccc|ccc|ccc|}
\hline

\multirow{2}{*}{\textbf{Task}} & \multirow{2}{*}{\textbf{Principle}}
& \multicolumn{3}{|c|}{\textbf{ChatGPT}}
& \multicolumn{3}{|c|}{\textbf{Gemini}}
& \multicolumn{3}{|c|}{\textbf{Claude}} \\
\cline{3-11}

& & P1 & P2 & P3 & P1 & P2 & P3 & P1 & P2 & P3\\

\hline
\multirow{3}{*}{\textbf{HMS}}
& Abstraction        & 0 & 0 & 0 & 0 & 0 & 0 & 0 & 0 & 0 \\ \cline{2-11}
& Encapsulation      & 1 & 1 & 1 & 1 & 1 & 1 & 1 & 1 & 1 \\ \cline{2-11}

\hline
\multirow{4}{*}{\textbf{Sensor Network}}
& Abstraction        & 0 & 0 & 0 & 0 & 0 & 0 & 0 & 0 & 0 \\ \cline{2-11}
& Encapsulation      & 0 & 0 & 0 & 1 & 1 & 1 & 1 & 1 & 1 \\ \cline{2-11}
\hline
\end{tabular}

\caption{M1.2: Principle Adherence Score — Rule-injected Baseline Averaged across 10 runs}
\label{tab:PASfair}
\end{table*}
\begin{table*}[t]
\centering
\scriptsize
\begin{tabular}{|c|c|ccc|ccc|ccc|}
\hline

\multirow{2}{*}{\textbf{Task}} & \multirow{2}{*}{\textbf{Principle}}
& \multicolumn{3}{|c|}{\textbf{ChatGPT}}
& \multicolumn{3}{|c|}{\textbf{Gemini}}
& \multicolumn{3}{|c|}{\textbf{Claude}} \\
\cline{3-11}

& & P1 & P2 & P3 & P1 & P2 & P3 & P1 & P2 & P3 \\

\hline
\multirow{3}{*}{\textbf{HMS}}
& Abstraction        & 0.5 & 0.5 & 0.5 & 0 & 0 & 0 & 0 & 0 & 0 \\ \cline{2-11}
& Encapsulation      & 1 & 1 & 1 & 1 & 1 & 1 & 1 & 1 & 1 \\ \cline{2-11}

\hline
\multirow{4}{*}{\textbf{Sensor Network}}
& Abstraction        & 0 & 0 & 0 & 1 & 1 & 1 & 0.5 & 0.5 & 0.5 \\ \cline{2-11}
& Encapsulation      & 1 & 1 & 1 & 1 & 1 & 1 & 1 & 1 & 1 \\ \cline{2-11}
\hline
\end{tabular}

\caption{M1.2: Principle Adherence Score — Preference-Based Approach Averaged across 10 runs}
\label{tab:PASpref}
\end{table*}

\begin{table*}[t]
\centering
\scriptsize

\begin{tabular}{|c|c|ccc|ccc|ccc|}
\hline

\multirow{2}{*}{\textbf{Task}} & \multirow{2}{*}{\textbf{Patterns}}
& \multicolumn{3}{|c|}{\textbf{ChatGPT}}
& \multicolumn{3}{|c|}{\textbf{Gemini}}
& \multicolumn{3}{|c|}{\textbf{Claude}}\\

\cline{3-11}

& & P1 & P2 & P3 & P1 & P2 & P3 & P1 & P2 & P3  \\

\hline
\multirow{3}{*}{\textbf{HMS}}
& Strategy Pattern   & 0 & 0 & 0 & 1 & 1 & 1 & 0 & 0 & 0\\
\hline
\multirow{4}{*}{\textbf{Sensor Netwrok}}
& Strategy Pattern   & 0 & 0 & 0 & 0 & 0 & 0 & 0 & 0 & 0 \\ \cline{2-11}
& Observer Pattern   & 0 & 0 & 0 & 0 & 0 & 0 & 0 & 0 & 0 \\

\hline
\end{tabular}

\caption{M1.3: Pattern Emergence Score — Standard Baseline Averaged across 10 runs}
\label{tab:PESstandard}
\end{table*}
\begin{table*}[t]
\centering
\scriptsize
\begin{tabular}{|c|c|ccc|ccc|ccc|}
\hline

\multirow{2}{*}{\textbf{Task}} & \multirow{2}{*}{\textbf{Patterns}}
& \multicolumn{3}{|c|}{\textbf{ChatGPT}}
& \multicolumn{3}{|c|}{\textbf{Gemini}}
& \multicolumn{3}{|c|}{\textbf{Claude}} \\
\cline{3-11}

& & P1 & P2 & P3 & P1 & P2 & P3 & P1 & P2 & P3\\

\hline
\multirow{3}{*}{\textbf{HMS}}
& Strategy Pattern   & 0 & 0 & 0 & 1 & 0 & 0 & 0 & 0 & 0 \\

\hline
\multirow{4}{*}{\textbf{Sensor Network}}
& Strategy Pattern   & 0 & 0 & 0 & 0 & 0 & 0 & 0 & 0 & 0 \\
\cline{2-11}
& Observer Pattern   & 0 & 0 & 0 & 0 & 0 & 0 & 0 & 0 & 0 \\

\hline
\end{tabular}

\caption{M1.3: Pattern Emergence Score — Rule-injected Baseline Averaged across 10 runs}
\label{tab:PESfair}
\end{table*}
\begin{table*}[t]
\centering
\scriptsize
\begin{tabular}{|c|c|ccc|ccc|ccc|}
\hline

\multirow{2}{*}{\textbf{Task}} & \multirow{2}{*}{\textbf{Principle}}
& \multicolumn{3}{|c|}{\textbf{ChatGPT}}
& \multicolumn{3}{|c|}{\textbf{Gemini}}
& \multicolumn{3}{|c|}{\textbf{Claude}} \\
\cline{3-11}

& & P1 & P2 & P3 & P1 & P2 & P3 & P1 & P2 & P3 \\

\hline
\multirow{3}{*}{\textbf{HMS}}
& Strategy Pattern   & 0.5 & 0.5 & 0.5 & 1 & 1 & 1 & 1 & 1 & 1 \\

\hline
\multirow{4}{*}{\textbf{Sensor Network}}

& Strategy Pattern   & 0.5 & 0.5 & 0.5 & 0 & 0 & 0 & 0 & 0 & 0\\ \cline{2-11}
& Observer Pattern   & 0 & 0 & 0 & 0 & 0 & 0 & 0 & 0 & 0\\

\hline
\end{tabular}

\caption{M1.3: Pattern Emergence Score — Preference-Based Approach Averaged across 10 runs}
\label{tab:PESpref}
\end{table*}

\subsection{Answer to RQ3}

A central contribution of this study is the identification of substantial reliability differences across frontier LLMs (see table \ref{tab:SI}). Claude produced varied architectures across approaches and paraphrased prompts but remained highly consistent across repeated runs for the same prompt. This behavior suggests strong decoding stability: once conditioned, the model tends to converge toward a similar structural solution. However, stability did not guarantee correctness, as several diagrams consistently omitted critical architectural relationships (see results from M1.1). GPT showed sensitivity to the modeling approach but was largely unaffected by paraphrasing or repeated executions. Once appropriate architectural scaffolding was established, the model behaved predictably, indicating a high degree of controllability. Importantly, GPT’s errors appeared systematic rather than stochastic, suggesting bounded reasoning limitations rather than instability. Gemini, in contrast, exhibited significant variability across both runs and prompts, pointing to elevated stochasticity in structural generation. Although semantic intent often remained similar, diagram topology frequently shifted — with fluctuating numbers of classes and relationships — indicating architectural drift rather than deliberate refinement (see tables \ref{tab:HMS-cc}-\ref{tab:WSRel}).

An important observation is that semantic consistency did not imply structural consistency. For architecture-centric tasks, structural variance is particularly consequential because inconsistencies may propagate into downstream implementation defects.

\textit{Key Insight:} Reliability is strongly model-dependent. Model choice may therefore be as critical as prompting strategy when deploying LLMs for architectural workflows.
\begin{table}
    \centering
    \scriptsize
    \begin{tabular}{|c|c|c|c|}
    \hline
    \textbf{Model} & \textbf{Standard Baseline} & \textbf{Rule Baseline} & \textbf{Preference Prompting}
    \\
    \hline
    \textbf{Claude} & 0.77 & 1 &1 \\ \hline
    \textbf{ChatGPT} & 1 & 1 &1 \\ \hline
    \textbf{Gemini} & 0.905 & 0.905 &0.907 \\ \hline

    \end{tabular}
    \caption{M1.3: Stability Index}
    \label{tab:SI}
\end{table}

\subsection{Answer to RQ4}

When task complexity increased from medium-scale problems to the high-complexity sensor network scenario, performance declined across all models — including under the preference-based setting. Most notably, no model successfully inferred the need for the Observer pattern in an event-driven architecture (see tables \ref{tab:PESstandard}-\ref{tab:PESpref}).

The Observer pattern requires recognizing implicit behavioral dependencies rather than explicit structural cues. The inability to detect this pattern suggests that current models rely heavily on pattern salience and struggle when architectural solutions must be inferred from dynamic system behavior.

This degradation indicates a ceiling effect in current design reasoning capabilities: as domains grow more complex and patterns become less obvious, models require stronger guidance or more advanced alignment mechanisms.

\textit{Key Insight:} Architectural inference — particularly for behavior-driven patterns — remains a major open challenge for LLM-based design systems.

\subsection{Cross-Cutting Implications}

Taken together, the results reframe LLM-based design generation as both a capability and a reliability problem. First, model-level behavioral characteristics strongly influence outcomes. The observed stability regimes suggest that architectural dependability cannot be assumed across frontier systems. Second, preference-based prompting shows clear promise as a mechanism for steering models toward principled designs. However, persistent design smells and incomplete constraint satisfaction indicate that alignment alone is not sufficient for fully reliable synthesis. Third, stability should be treated as a first-class evaluation criterion. A model that produces a correct architecture once but cannot reproduce it consistently presents operational risks in professional development settings.

The findings suggest that the central question is no longer whether LLMs can generate UML diagrams, but whether they can function as dependable architectural collaborators. While preference-based methods improve the likelihood of design-oriented outputs, substantial gaps remain in structural reasoning, pattern inference, and behavioral reliability.

\section{Limitations}
This study is preliminary and limited to two domains, and single modality generation tasks; design principles used as benchmark for measuring diagram quality; relies on expert labeling; and specific prompt design. It evaluates LLMs on design task using a small synthetic preference dataset before implementing a computationally costly, full RLHF-like pipeline. A broader validation of proposed approach on various domains, and in multi-artifact setup is left to future work.
\section{Conclusion and Future Work}
This study systematically investigated the capability and reliability of Large Language Models (LLMs) in generating UML class diagrams from domain-only descriptions. We highlighted the distinction between diagram translation and design synthesis, emphasizing that effective software design requires reasoning about latent constraints, abstractions, and architectural patterns. Through 540 controlled experiments across three LLMs, three prompting strategies, and multiple prompt paraphrases, we observed that model-level behavior significantly influences design reliability. Preference-based few-shot alignment improved adherence to design principles and pattern-consistent structures, but non-determinism persisted, particularly in complex domains.

The findings of this study suggest a more consequential question: can models behave like dependable architectural collaborators?
The findings further suggest that achieving dependable LLM-assisted software design is not solely a matter of prompting strategy; careful selection of the model and robustness-aware design methods are equally critical, and behavioral stability should be considered a core evaluation criterion alongside syntactic correctness and pattern adherence.

For future work, we plan to extend this study in several directions: investigating hybrid approaches that combine preference-based prompting with explicit architectural reasoning modules to further improve design inference, expanding benchmarks to include additional domain complexities, exploring techniques to reduce non-determinism in LLM outputs, including model fine-tuning, constrained decoding, and ensemble approaches.

\section{Acknowledgment}
The authors acknowledge the use of GenerativeAI tools, specifically ChatGPT for non-substantive editorial assistance in refining the manuscript’s language, grammar, and structural flow.

\end{document}